# Accurate Modeling of Interfacial Thermal Transport in van der Waals Heterostructures via Hybrid Machine Learning and Registry-Dependent Potentials

Wenwu Jiang,[1,#] Hekai Bu,[1,#] Ting Liang,[2,#] Penghua Ying,[3] Zheyong Fan,[4] Jianbin Xu,[2] Wengen Ouyang[1,5*]

[1]*Department of Engineering Mechanics, School of Civil Engineering, Wuhan University, Wuhan, Hubei 430072, China*

[2]*Department of Electronic Engineering and Materials Science and Technology Research Center, The Chinese University of Hong Kong, Shatin, N.T., Hong Kong SAR, 999077, P. R. China*

[3]*Department of Physical Chemistry, School of Chemistry, Tel Aviv University, Tel Aviv 6997801, Israel*

[4]*College of Physical Science and Technology, Bohai University, Jinzhou 121013, China*

[5]*State Key Laboratory of Water Resources Engineering and Management, Wuhan University, Wuhan, Hubei 430072, China*

[#]These authors contribute equally to this work.

[*]Corresponding authors. Email: w.g.ouyang@whu.edu.cn

**ABSTRACT**

Two-dimensional transition metal dichalcogenides (TMDs) exhibit remarkable thermal anisotropy due to their strong intralayer covalent bonding and weak interlayer van der Waals (vdW) interactions. However, accurately modeling their thermal transport properties remains a significant challenge, primarily due to the computational limitations of density functional theory (DFT) and the inaccuracies of classical force fields in non-equilibrium regimes. To address this, we use a recently developed hybrid computational framework that combines machine learning potential (MLP) for intralayer interactions with registry-dependent interlayer potential (ILP) for anisotropic vdW interlayer interaction, achieving near quantum mechanical accuracy. This approach demonstrates exceptional agreement with DFT calculations and experimental data for TMD systems, accurately predicting key properties such as lattice constants, bulk modulus, moiré reconstruction, phonon spectra, and thermal conductivities. The scalability of this method enables accurate simulations of TMD heterostructures with large-scale moiré superlattices, making it a transformative tool for the design of TMD-based thermal metamaterials and devices, bridging the gap between accuracy and computational efficiency.

# 1. INTRODUCTION

In recent years, two-dimensional (2D) transition metal dichalcogenides (TMDs) have received significant research attention due to their outstanding electrical[1-3], thermal[4-8], tribological[9-12], and optical properties[13, 14]. These attributes are closely associated with their highly anisotropic nature, characterized by strong intralayer bonded interactions and weak interlayer van der Waals (vdW) interactions. For instance, the randomly twisted TMD structures show ultra-high in-plane and ultra-low cross-plane thermal conductivities, respectively, resulting in an exceptionally thermal conductivity anisotropy ratio[6]. To understand the microscopic mechanism, an accurate description of intralayer and interlayer interactions of 2D TMDs is essential. Typically, the solution of the phonon Boltzmann transport equation (BTE) within the framework of density functional theory (DFT)[15, 16], serves as the primary accurate method to explore the thermal transport properties of 2D materials. However, inherent difficulties associated with the solution of the BTE on short time and length scales have significantly limited the consideration of non-Fourier effects in practically important heat conduction problems[17]. Moreover, the computational burden of DFT for the interatomic force constants constrains its application when investigating the thermal transport properties in large-scale atomistic systems or complex interface structures[18]. In these situations, molecular dynamics (MD) simulations based on a force field approach emerge as a more efficient alternative, however, their accuracy is highly dependent on the adopted force fields.

Currently, a variety of empirical many-body force fields have been developed to model the intralayer interaction of monolayer TMDs, including Stillinger-Weber (SW) potential[19, 20], and reactive empirical bond-order (REBO) potential[21, 22]. While these force fields provide a good description of mechanical and thermodynamic properties of monolayer TMDs at the equilibrium regime, they are inadequate for modeling the behavior of defective or edged TMDs under non-equilibrium conditions[23]. For multilayer TMD systems, a pairwise Lennard-Jones (LJ) potential is commonly used to describe the long-range vdW interactions between layers. Combined with the intralayer potential for monolayer TMDs, this method offers an efficient approach to simulate the physical properties of multilayer TMDs. However, recent studies show that the phonon spectra calculated with this approach deviate significantly from the experimental observations[24-26]. Accurate characterization of phonon properties is crucial for depicting thermal transport behaviors in TMDs.

Recently, machine learning potentials (MLPs)[27-29], have been developed to capture the interatomic interactions of 2D materials with quantum-mechanical accuracy, enabling more reliable predictions of complex material behavior. However, modeling long-range vdW interactions using MLPs presents a significant challenge in balancing physical accuracy with computational efficiency. The scalability of local models depends on limiting the number of interactions that need to be evaluated, while long-range interactions (electrostatic interactions, charge transfer, and dispersion effects) inherently involve a broader range of interactions, posing a challenge for efficiently capture within a local model framework[30-32]. Although MLP models with sufficiently large cutoff ranges are theoretically capable of capturing long-range vdW interactions, achieving an acceptable level of accuracy typically requires extensive cutoff testing and imposes substantial computational costs[33]. This issue is particularly pronounced given that the energy scales of vdW interactions (30–60 meV/atom for graphite and 13-22 meV/atom for $MoS_2$)[24, 34] are significantly smaller than those of intralayer interactions (~7.3 eV/atom for graphite and ~4.5 eV/atom for $MoS_2$)[34, 35], necessitating disproportionately large datasets to adequately explore the vast configurational space of vdW interactions. As such, merely increasing the size of the local environment to account for long-range interactions is not a viable solution. To effectively model long-range interactions, an optimal approach would involve adopting a multiscale feature representation, rather than relying solely on local models. This strategy would facilitate a better balance between computational efficiency and physical accuracy, thereby addressing the limitations of current MLP models in capturing long-range vdW interactions in 2D materials[32, 33, 36].

To tackle this challenge, various approaches have been explored to enhance MLPs for better incorporation of dispersion interactions. For instance, Muhli et al.[37] refined the approach by determining the dispersion coefficient and damping function using a local descriptor to further improve computational accuracy. Wen et al.[36] introduced an additional attractive term into MLPs, which depends on the interatomic distance and a fitting parameter, modulated by specific switching/damping functions with four additional fitting parameters. This method demonstrated high accuracy in describing binding and sliding energies in bilayer graphene. Ying et al.[33] combined the D3 dispersion correction from DFT with the neuroevolution potential (NEP) for accurately describing the binding and sliding energies of bilayer graphene. While these approaches have significantly enhanced the performance of MLPs through dispersion corrections, challenges persist, particularly when extending their applicability to diverse LEGO-like vdW heterostructures[38].

Our recent studies have shown that registry-dependent interlayer potentials (ILPs) provide an accurate description of anisotropic interlayer vdW interactions. These ILPs effectively model the mechanical[39, 40], thermal transport[25, 26], and tribological properties[41-51] of various vdW heterostructures, including graphene/hexagonal boron nitride (*h*-BN)[47, 52], TMDs[24, 53], and others[54-59]. Based on these findings, we recently proposed a hybrid computation framework that integrates MLP with ILP for a more accurate description of layered vdW interfaces including graphene, *h*-BN, MoS$_2$ and their heterojunctions[38]. In the present study, we extended this approach to the TMD homogeneous and heterogeneous structures, in which MLP and ILP are employed to describe intralayer and interlayer interactions in these structures, respectively. This combined approach significantly reduces the required training configurations while maintaining computational efficiency and high accuracy for describing both intralayer and interlayer interaction of TMD system. To get a better balance between high efficiency and sufficient accuracy, we choose the NEP[60-62] model to characterize the intralayer interactions of 2D TMD materials. Moreover, we have additionally evaluated the applicability of the NEP method combined with D3 dispersion correction to the MoS$_2$ system.

The MD simulations employing the NEP+ILP approach demonstrate good alignment with DFT results and experimental data across various calculations, including the calculations of intra- and interlayer lattice constants, bulk modulus, atomic reconstruction in moiré superlattices, phonon spectra, as well as in-plane and cross-plane thermal conductivities. The MD simulations for lattice constants, bulk modulus, phonon dispersion, and atomic reconstruction calculations were performed using the large-scale atomic/molecular massively parallel simulator (LAMMPS)[63] package. The calculations for in-plane and cross-plane thermal conductivities were performed using the homogeneous nonequilibrium molecular dynamics (HNEMD)[64] method from the Graphics Processing Units Molecular Dynamics (GPUMD)[65] simulation package.

## 2. METHODS

### 2.1. Computational Framework via integrating ILP with NEP

The functional form of the NEP+ILP approach has been recently introduced[38], nevertheless, for the sake of completeness, we briefly reiterate this definition here. In this hybrid computational framework, the total potential energy of the system consists of intralayer and interlayer energies. Here, we employed NEP model and ILP to describe intralayer and interlayer interactions of TMD structures, respectively (see **Fig. 1**). The total potential energy ($E_{\text{tot}}^{\text{NEP+ILP}}$) of a multilayer system thus can be expressed as:

$$E_{\text{tot}}^{\text{NEP+ILP}} = \sum_{k=1}^{M}\sum_{l=1}^{M}\left[E_{kl}^{\text{NEP}}\delta_{kl} + E_{kl}^{\text{ILP}}(1-\delta_{kl})\right], \tag{1}$$

where $E_{kl}^{\text{NEP}}$ and $E_{kl}^{\text{ILP}}$ represent the potential energy of the NEP model and the ILP model between the $k^{\text{th}}$ and $l^{\text{th}}$ layer, respectively. $M$ is the total number of layers of the system. $\delta_{kl}$ is the Kronecker delta function.

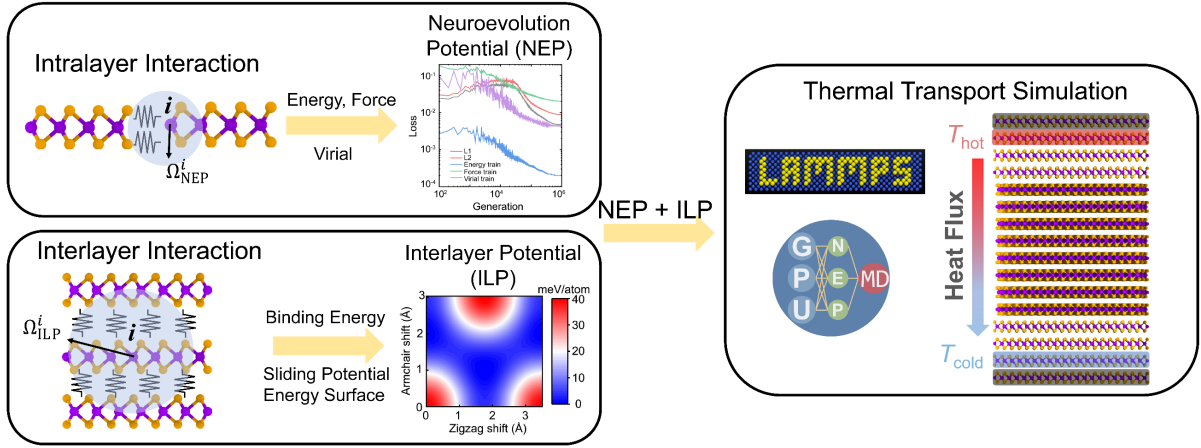

**Fig. 1 | The methodology for integrating anisotropic interlayer potential with machine learning potential.** The approach involves parameterizing the NEP to model the strong chemical interactions within the TMD layer and ILP to capture weak vdW interactions between TMD layers, respectively. Once developed, this hybrid force field is utilized to simulate and predict thermal transport and mechanical properties of TMD heterostructures through the integration with the LAMMPS[63] and the GPUMD[65] simulation packages.

The intralayer chemical bonded interaction of within a monolayer TMD is modeled via the NEP, which is based on neural networks (NN) and trained using separable natural evolution strategies (SNES). The NEP model is currently a widely used MLP approach in mechanics and thermal transport studies[66-68]. Because the NEP solely describes intralayer interactions, we only need to train the NEP for a monolayer system and then directly apply it to multilayer systems in conjunction with the ILP. This approach, by decoupling short-range intralayer and long-range interlayer interactions, significantly reduces the requirement for reference datasets while enhancing both the accuracy and transferability of potential models for complex van der Waals heterostructures[38]. Following the standard Behler-Parrinello high-dimensional NN potential approach[69], the site energy of a single layer (denoted as $E_{kk}^{\text{NEP}}, k = 1,2,\cdots$) can be written as follows[61]:

$$E_{kl}^{\text{NEP}}\delta_{kl} = E_{kk}^{\text{NEP}} = \sum_{i\in\Omega^k}^{N}\left[\sum_{\mu=1}^{N_{\text{neu}}} w_\mu^{(1)}\tanh\left(\sum_{\nu=1}^{N_{\text{des}}} w_{\mu\nu}^{(0)} q_\nu^i - b_\mu^{(0)}\right) - b^{(1)}\right], \quad (2)$$

where $\Omega^k$ is the set of atoms in the $k^{th}$ layer, $N$ is total number of atoms in the layer, $N_{neu}$ is the number of neurons in hidden layer, and $N_{des}$ is the number of descriptors. The $\tanh(\cdot)$ is the activation function, and $w_{\mu\nu}^{(0)}$, $w_{\mu}^{(1)}$, $b_{\mu}^{(0)}$, and $b^{(1)}$ are trainable weight and bias parameters in the NN. The $q_{\nu}^{i}$ represents the local atomic environment descriptor, which consists of several radial and angular components. The radial descriptor components (with a cutoff radius $r_c^R$) are constructed based on Chebyshev polynomials, while the angular descriptor components (with a cutoff radius $r_c^A$), which depend on angular information, are constructed based on spherical harmonics similar to the atomic cluster expansion (ACE) method[70]. The radial descriptor components $q_n^i (0 \leq n \leq n_{max}^R)$ are constructed as

$$q_n^i = \sum_{j \in \Omega_i^{NEP}} g_n(r_{ij}), \tag{3}$$

where $r_{ij}$ is the distance between atom $i$ and atom $j$, $g_n$ is the $n^{th}$ expansion radial descriptor function. The atoms included in the neighbor list $\Omega_i^{NEP}$ of atom $i$ are identified within the same layer, using a cutoff radius $R_{cut}^{NEP} = r_c^{R/A}$. The angular descriptor components, considered three-body terms in NEP, can be expressed as:

$$q_{nl}^i = \sum \sum_{j,k \in \Omega_i^{NEP}} g_n(r_{ij}) g_n(r_{ik}) P_l(\theta_{ijk}), \tag{4}$$

where $P_l(\cdot)$ is the Legendre polynomial of order $l$, and $\theta_{ijk}$ is the angle for the triplet $(ijk)$ with atom $i$ in the centra.

The interlayer interaction of the TMD system is modeled by the registry-dependent ILP. Because the ILP solely describes interlayer interactions, we only need to train the ILP for bilayer systems and then directly apply it to multilayer systems in conjunction with the NEP. The structure of the interlayer potential consists of two interactions, which are short-range repulsion and long-range attraction, as shown in the following expressions:

$$E_{kl}^{ILP}(1 - \delta_{kl}) = \sum_{i \in \Omega^k} \sum_{j \in \Omega^l \cap \Omega_i^{ILP}}^{l \neq k} \text{Tap}(r_{ij}) [V_{rep}(r_{ij}, \boldsymbol{n}_i, \boldsymbol{n}_j) + V_{att}(r_{ij})], \tag{5}$$

where $\Omega^k$ and $\Omega^l$ are the set of atoms in the $k^{th}$ and $l^{th}$ layer, respectively, and

$$\text{Tap}(r_{ij}) = 20 \left(\frac{r_{ij}}{R_{cut}}\right)^7 - 70 \left(\frac{r_{ij}}{R_{cut}}\right)^6 + 84 \left(\frac{r_{ij}}{R_{cut}}\right)^5 - 35 \left(\frac{r_{ij}}{R_{cut}}\right)^4 + 1 \tag{6}$$

provides a continuous long-range cutoff (up to the third derivative) which used to dampen the interaction in adjacent layers at interatomic separations larger than $R_{cut}^{ILP} = 16$ Å. The $r_{ij}$ is the Euclidean distance between atom $i$ in layer $k$ and atom $j$ in layer $l$, where atom $j$ belongs to the neighbor list $\Omega_i^{ILP}$ of atom $i$ (see **Fig. 1**). These neighbors are identified in adjacent layers using a cutoff radius of $R_{cut}^{ILP}$. The short-range repulsion and long-range attraction are evaluated using the following pairwise expressions:

$$V_{\text{att}}(r_{ij}) = -\frac{1}{1+e^{-d_{ij}[r_{ij}/(s_{R,ij} \cdot r_{ij}^{\text{eff}})-1]}} \frac{C_{6,ij}}{r_{ij}^6}, \tag{7}$$

$$V_{\text{rep}}(r_{ij}, \boldsymbol{n}_i, \boldsymbol{n}_j) = e^{\alpha_{ij}\left(1-\frac{r_{ij}}{\beta_{ij}}\right)} \left\{\varepsilon_{ij} + C_{ij}\left[e^{-(\rho_{ij}/\gamma_{ij})^2} + e^{-(\rho_{ji}/\gamma_{ij})^2}\right]\right\}, \tag{8}$$

where $C_{6,ij}$ is the pairwise dispersion coefficient, $r_{ij}^{\text{eff}}$ is the sum of the effective equilibrium vdW atomic radii, and $d_{ij}$ and $s_{R,ij}$ are unit-less parameters defining the steepness and onset of the short-range Fermi−Dirac type damping function. Moreover, $\varepsilon_{ij}$ and $C_{ij}$ are constants that set the energy scales associated with the isotropic and anisotropic repulsions, respectively, $\beta_{ij}$ and $\gamma_{ij}$ set the corresponding interaction ranges, and $\alpha_{ij}$ is a parameter that sets the steepness of the isotropic repulsion function. $\boldsymbol{n}_i$ and $\boldsymbol{n}_j$ are the corresponding local normal vectors of atoms $i$ and $j$. $\rho_{ij}(\rho_{ji})$ is the lateral distances of atom $j(i)$ to the surface normal, $\boldsymbol{n}_i(\boldsymbol{n}_j)$ of atom $i(j)$:

$$\begin{cases} \rho_{ij}^2 = r_{ij}^2 - (\boldsymbol{r}_{ij} \cdot \boldsymbol{n}_i)^2 \\ \rho_{ji}^2 = r_{ji}^2 - (\boldsymbol{r}_{ji} \cdot \boldsymbol{n}_j)^2 \end{cases}. \tag{9}$$

*2.2. Neuroevolution Potential (NEP) Model Training for TMDs*

In this work, the cutoff radius of radial and angular descriptor parts are $r_c^R = 5$ Å and $r_c^A = 5$ Å used for NEP model training, respectively. For both the radial and angular descriptor components, we employ 9 radial functions, each one being a linear combination of 13 Chebyshev polynomials. Following the ACE approach for constructing the angular descriptor components, we use three and four-body correlations in the spherical harmonics up to degree *l* = 4 and *l* = 2, respectively. For all six MX$_2$ (where M = Mo/W and X = S/Se/Te) systems, we used the same hyperparameters. More detailed hyperparameters regarding the NEP model training are presented in Section 1 of Supporting Information.

As mentioned above, the NEP dataset only need to include monolayer configurations, showcasing significant data savings. We first generated 50 structures by applying random cell deformations (-3 to 3%) and atomic displacements (within 0.1 Å) starting from the optimized structure. Subsequently, to obtain configurations with actual temperature fluctuations, we conduct empirical potential-driven MD simulations (use SW potential specifically for TMD systems) under constant volume conditions at target temperatures of 300, 600, and 900 K, sampling 200 structures. For the NEP dataset, we have a total of 250 frames for each MX$_2$ system.

To obtain the energy, force, and virial data for NEP training, we performed DFT calculations using the Perdew-Burke-Ernzerhof functional[71] and the projector-augmented wave method that implemented in the Vienna ab initio simulation package (VASP)[72, 73]. For all considered configurations, the vaccume size along the *c*-axis of TMD systems is set to 50 Å to avoid spurious interactions between neighboring images. The energy convergence threshold for the electronic self-consistent loop is set as $10^{-8}$ eV, with an energy cutoff of 850 eV, utilizing the projector-augmented wave (PAW) method[74]. We sampled the Brillouin zone using a dense Γ-centered grid with a K-point density of 0.15/Å. The tetrahedron method was utilized to calculate the total energy.

*2.3. The Interlayer Potential (ILP) Parameterization for TMDs*

The ILP provides a successful description of the interlayer interactions in various layered materials, such as graphene, *h*-BN, and TMD-based systems[24, 47, 52-59], and also predicts reliable thermal transport, mechanical, and tribological properties of those layered materials[25, 26, 39-51]. The function of the parameterized ILP in this work follows the same form as that used previously for TMD system[24, 53]. In our previous works focused on TMDs[24, 53], we successfully developed and benchmarked the ILP for $MX_2$ systems (where M = Mo/W and X = S/Se) using the Heyd-Scuseria-Ernzerhof (HSE) hybrid density functional approximation, in conjunction with non-local many-body dispersion (MBD-NL)[75] long range corrections. This thorough benchmarking of the ILP confirmed the robustness of the DFT method (HSE+MBD-NL), as evidenced by the alignment of the simulation outcomes with an array of experimental data. Unfortunately, the utilization of this DFT method (HSE+MBD-NL) is not viable for transition-metal telluride (Te) due to the excessive computational burden associated with the all-electron and HSE calculations. As a solution, we turn to the Perdew-Burke-Ernzerhof (PBE) density functional approximation instead to generate the DFT reference data for structures containing transition-metal telluride. Given the small disparity between the DFT energies obtained through PBE+MBD-NL and HSE+MBD-NL methods for $MoTe_2$ and $WTe_2$, the accuracy of our adopted PBE+MBD-NL approach for $MoTe_2$ and $WTe_2$ remains satisfactory (see Section 2 of Supporting Information). For each homojunctions $MoTe_2$ and $WTe_2$ system, we computed five binding energy (BE) curves and two sliding potential energy surfaces (PESs). Each BE curve and sliding PES contain 31 and 132 data points, respectively.

## 3. BENCHMARK TESTS

*3.1. Intra- and Interlayer Lattice Constants*

To evaluate the structural performance of TMD systems under hydrostatic pressure using the NEP+ILP hybrid computational framework, we conducted MD simulations to calculate the structural parameters (including intralayer lattice constant $a_0$, interlayer lattice constant $c_0$, and unit-cell volume $V$) of bulk $MoS_2$, $MoSe_2$, $MoTe_2$, $WS_2$, $WSe_2$, and $WTe_2$. We further analyzed their dependence on hydrostatic pressure ($P$) and compared the results with existing experimental data and DFT results. All MD simulations in this study are performed with the LAMMPS[63]. Herein, the NEP potential is employed to describe the intralayer interactions of TMD structures. The interlayer interactions are modeled via our developed anisotropic ILP. The model system for calculating the structural parameters of bulk homojunctions consists of 12 roughly rectangle layers, each containing 2500 chalcogenide atoms and 1250 transition metal atoms, and the adjacent layers are arranged in AA′ stacking modes. In the simulations, periodic boundary conditions are applied in all directions, and the time step for propagating the equations of motion is set to 1 fs. A Nose-Hoover thermostat with a time constant of 0.25 ps and a Nose-Hoover barostat with a time constant of 1.0 ps is used to maintain the system at a specified temperature and hydrostatic pressure, respectively. The entire system is equilibrated using the above NPT ensemble at a temperature of $T$ = 300 K and zero pressure for 200 ps. Then lattice parameters and volume are calculated by averaging over the last 100 ps. To extract the bulk modulus, we carried out a series of simulations for various pressures ranging from 0 to 14 GPa.

As shown in the middle and right columns of **Fig. 2**, both $a_0$ (intralayer lattice constant) and $c_0$ (interlayer lattice constant) parameters of bulk TMDs fall within the range of DFT data and experimental values. Particularly, the $a_0$ parameters of bulk $MoS_2$ (3.18 Å), $MoSe_2$ (3.33 Å), $MoTe_2$ (3.57 Å), $WS_2$ (3.19 Å), $WSe_2$ (3.33 Å), and $WTe_2$ (3.57 Å) show good agreement with experimental results and most of the dispersion-corrected DFT values, with deviations of 0.01, 0.02, 0.02, 0.01, 0.02, and 0.02 Å from experiments and DFT data (averaged over available measured data), respectively. Similarly, the $c_0$ parameters of bulk $MoS_2$ (12.56 Å), $MoSe_2$ (13.23 Å), $MoTe_2$ (14.29 Å), $WS_2$ (12.75 Å), $WSe_2$ (13.41 Å), and $WTe_2$ (14.34 Å) also show good agreement with both experimental results and most of the dispersion-corrected DFT values, with deviations of 0.02, 0.06, 0.04, 0.13, 0.17, and 0.02 Å from experiments and DFT data (averaged over available measured data), respectively. These results collectively indicate the validity range of the NEP+ILP force field for TMD systems.

*3.2. Bulk Modulus*

To further evaluate the performance of the NEP+ILP force field, we calculated the bulk modulus of TMD systems and compared these results with reference data. The pressure-volume (*P-V*) curves obtained from the MD simulations (detailed in Section 3.1 of Supporting Information) were fitted using three commonly employed equations of state (EOS) (i.e., the Murnaghan[76, 77] equation, the Birch-Murnaghan[78, 79] equation, and the Vinet[80, 81] equation) to extract the bulk modulus of the system. Detailed procedures are provided in Section 3 of Supporting Information. The Murnaghan bulk modulus of $MoS_2$, $MoSe_2$, $MoTe_2$, $WS_2$, $WSe_2$, and $WTe_2$ are 35.1 ± 5.8, 29.7 ± 1.6, 24.5 ± 1.0, 25.9 ± 4.2, 29.8 ± 1.6, and 23.8 ± 1.1 GPa, respectively. Similar values were obtained using the other EOS (see **Table S3** of Section 3 in Supporting Information). These values of bulk $MoS_2$, $MoSe_2$, $MoTe_2$, $WS_2$, and $WSe_2$ underestimate the experimental results by ∼ 30.5%, 44.8%, 42.3%, 58.2%, and 58.6%, respectively. Regarding bulk $WTe_2$, lacking available experimental reference data hampers definitive conclusions. Nonetheless, the calculated bulk modulus *B* from MD simulations (see left column of **Fig. 2**), utilizing all three EOS, falls within the range of DFT results. Overall, our findings indicate that the NEP+ILP force field approach offers a reasonable level of accuracy and reliability in characterizing the bulk properties of TMD systems across varying low- and high-pressure external conditions.

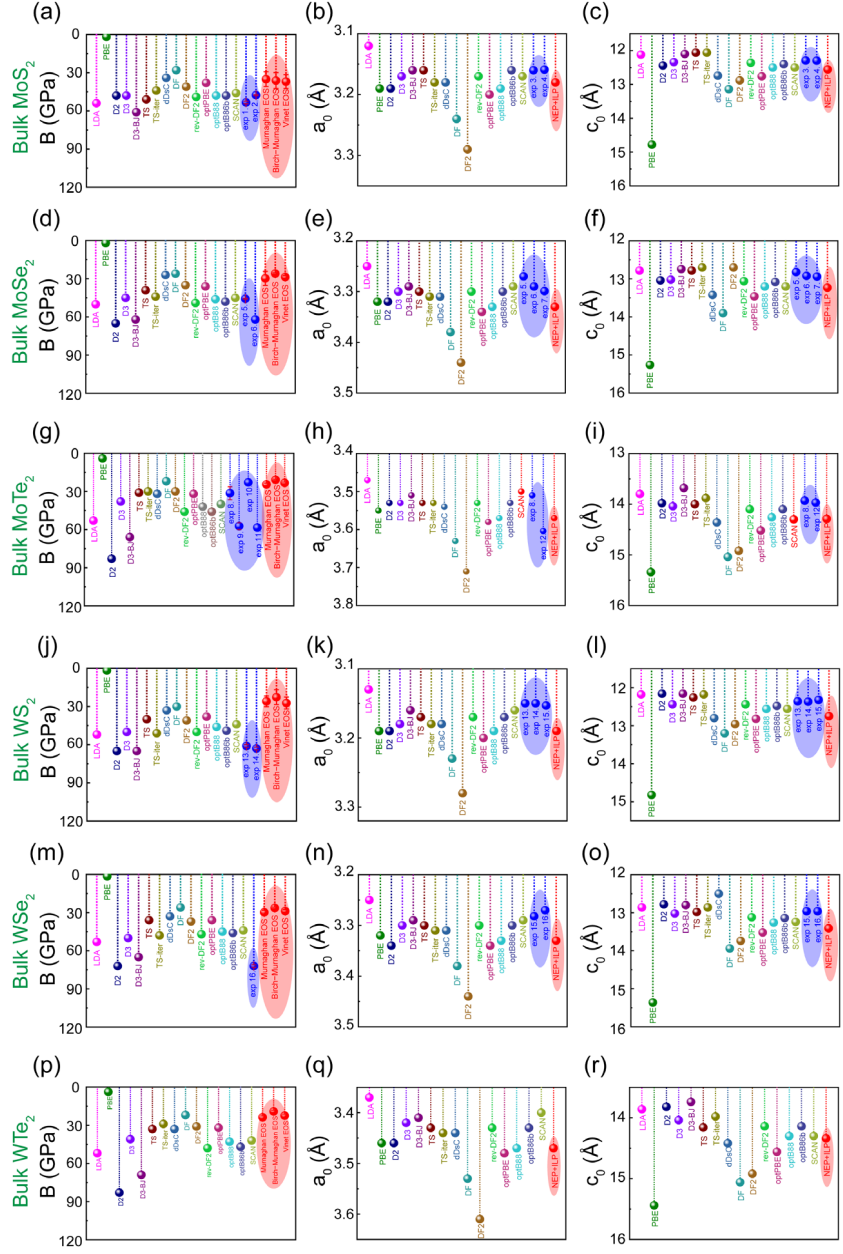

**Fig. 2 | Comparison of NEP+ILP calculations with DFT data and experimental results.** Bulk modulus (left column), $a_0$ (middle column), and $c_0$ (right column) were computed using NEP+ILP with available experimental and DFT data for AA′-stacked bulk TMD system ($MoS_2$, $MoSe_2$, $MoTe_2$, $WS_2$, $WSe_2$, and $WTe_2$). Reported experimental values and our MD simulation results are presented as blue and red circles, respectively. The DFT reference data are sourced from Ref. 82. The experimental reference data are extracted from following literatures: Ref. 83 (Exp. 1), Ref. 84 (Exp. 2), Ref.85 (Exp. 3, Exp. 7, and Exp. 12), Ref. 86 (Exp. 4), Ref. 87 (Exp. 5), Ref. 88 (Exp. 6), Ref. 89 (Exp. 8 and Exp. 9), Ref. 90 (Exp. 10 and Exp. 11), Ref. 91 (Exp. 13), Ref. 92 (Exp. 14), Ref. 93 (Exp. 15), and Ref. 94 (Exp. 16), respectively.

*3.3. Atomic Reconstruction in Twisted TMD Bilayers*

To demonstrate the applicability of the NEP+ILP force field for the description of twisted TMD interfaces, we evaluate its ability to capture the complex reconstructed moiré superstructures exhibited by twisted TMD interfaces. Here, we construct laterally periodic rectangular structures of twisted TMD bilayers at both the parallel and anti-parallel configurations, and perform geometry optimization using the NEP+ILP force field to describe the intra- and interlayer interactions, respectively.

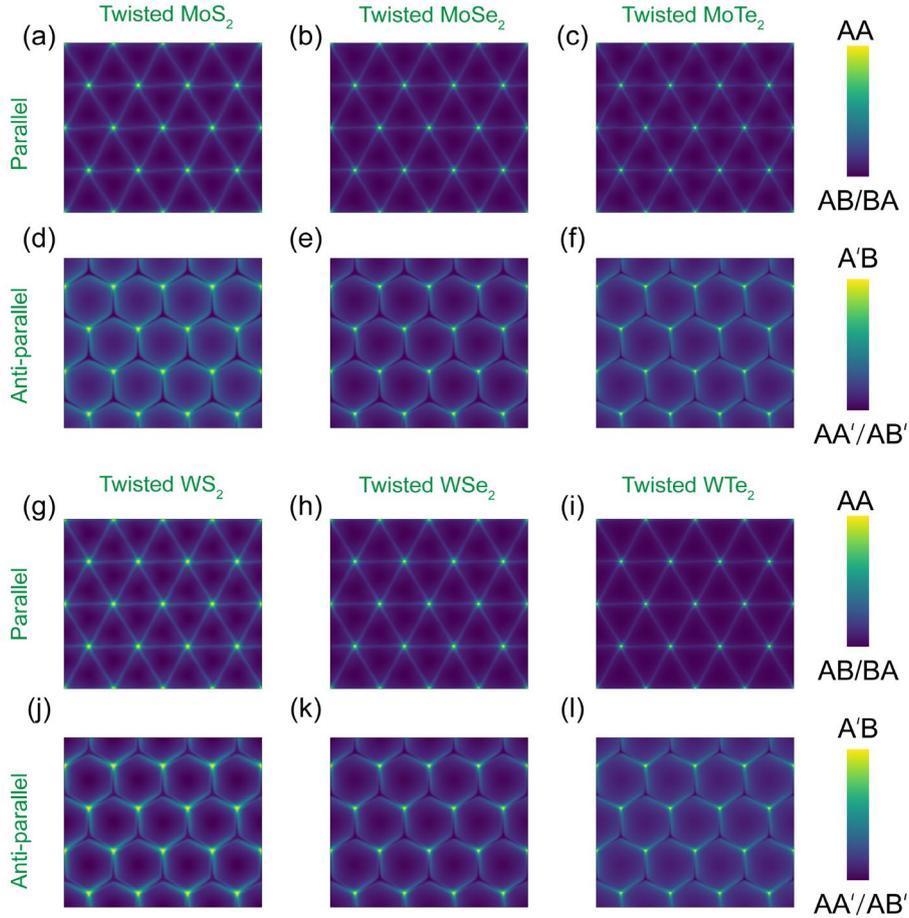

**Fig. 3 | Local registry index (LRI)[95, 96] mapping of TMD twisted interface.** LRI distribution of twisted MoS$_2$ (a,d), MoSe$_2$ (b,e), MoTe$_2$ (c,f), WS$_2$ (g,j), WSe$_2$ (h,k), and WTe$_2$ (i,l) bilayers in the parallel configuration (a-c and g-i) and the anti-parallel configuration (d-f and j-l) with $\theta$ = 0.25°, relaxed using the NEP+ILP force field. The yellow (black) markings stand for AA (AB and BA) stacks in parallel configuration and the A′B (AA′ and AB′) stacks in anti-parallel configuration, respectively. Here, we extend the registry index approach to quantify the interlayer commensurability of homogeneous interfaces of MoTe$_2$ and WTe$_2$ (see Section 4 of Supporting Information).

The geometry optimization performed follows our previous work[53], which ensures sufficient relaxation of residual stresses in twisted TMDs. As shown in **Fig. 3**, the triangular and hexagonal moiré superlattice patterns are clearly exhibited in parallel and anti-parallel twisted TMD bilayers with $\theta = 0.25°$, respecitvley. These patterns closely match the atomic reconstruction patterns observed experimentally in twisted homo-bilayers of TMDs[97, 98]. These results strongly affirm the validity of the parametrized ILP used in conjunction with machine learning force field in describing atomic reconstruction of moiré superlattices in twisted TMD interfaces.

*3.4. Phonon Spectra*

Phonon spectra are crucial for understanding the mechanical and thermal transport properties of materials. Here, we calculated the phonon spectra of monolayer and bulk homojunctions TMDs ($MoS_2$, $MoSe_2$, $MoTe_2$, $WS_2$, $WSe_2$, and $WTe_2$) using MD simulations performed with the LAMMPS package and compared them with the reference data to further assess the efficacy of the NEP+ILP force field. Due to the lack of experimental reference data (except for $MoS_2$), we conducted additional DFT calculations to obtain the phonon spectra of monolayer and bulk TMDs at the level of PBE+MBD[99, 100]. Specifically, the phonon spectra were calculated with a finite-difference method by using the phonopy package[101], which applied a displacement of 0.001 Å to each atom in three space directions and computed the dynamical matrix with PBE+MBD approaches via the VASP[72, 73]. Here, all the calculated bulk configurations are AA′ stacked at their equilibrium interlayer distances. The phonon spectra of monolayer $MoS_2$, $MoSe_2$, $MoTe_2$, $WS_2$, $WSe_2$, and $WTe_2$ calculated using NEP potential (red solid lines) are demonstrated in **Fig. 4**, showing good agreement with the DFT results (blue dashed lines) and exhibiting exceptional performance in characterizing both low- and high-frequency lattice vibrations.

**Figure 5** shows the phonon spectra of bulk $MoS_2$, $MoSe_2$, $MoTe_2$, $WS_2$, $WSe_2$, and $WTe_2$ with calculations performed using the NEP+ILP force field (red solid lines) and PBE+MBD (blue dashed lines). A noteworthy observation is that the NEP+ILP slightly underestimates the phonon energy branches along the A→Γ→M→K→Γ pathway compared to DFT data and experimental results. The minor disparities mentioned above primarily stem from the interlayer potential terms. Overall, the results of the low- and high energy out-of-plane branches are well described by the NEP+ILP force field.

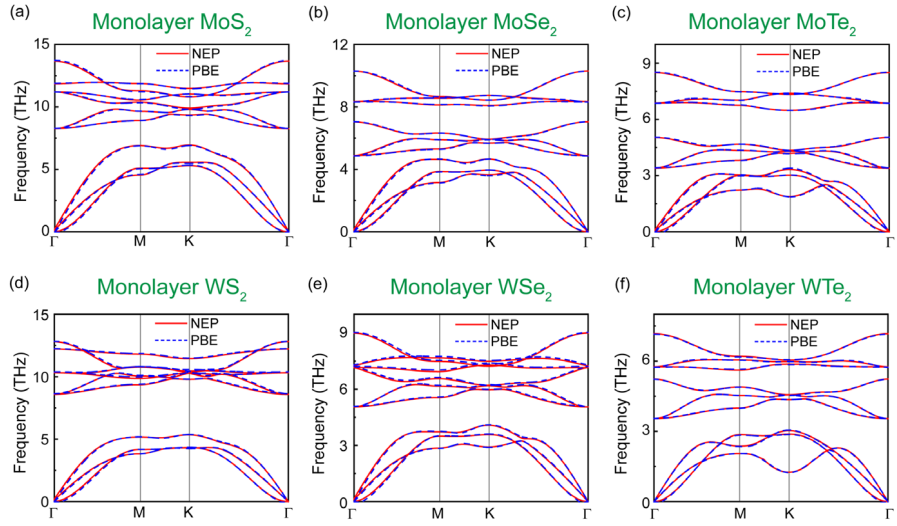

**Fig. 4 | Phonon spectra of monolayer TMD system.** Red solid lines and blue dashed lines are dispersion curves calculated using NEP+ILP force field and PBE for MoS$_2$ (a), MoSe$_2$ (b), MoTe$_2$ (c), WS$_2$ (d), WSe$_2$ (e), and WTe$_2$ (f), respectively.

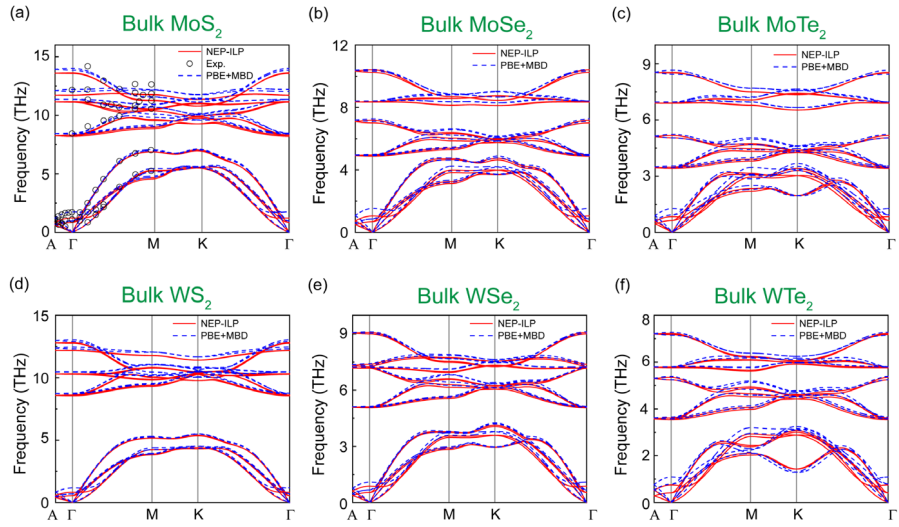

**Fig. 5 | Phonon spectra of bulk TMD system.** Red solid lines and blue dashed lines are dispersion curves calculated using NEP+ILP force field and PBE+MBD for MoS$_2$ (a), MoSe$_2$ (b), MoTe$_2$ (c), WS$_2$ (d), WSe$_2$ (e), and WTe$_2$ (f), respectively. Black circles present the experimental results of bulk MoS$_2$[102].

*3.5. In-plane Thermal Conductivity*

After confirming the reliability of NEP+ILP force field in describing phonon dispersion of TMD system, we applied them to calculate the in-plane thermal conductivity ($\kappa_{IP}$) using the HNEMD method[64]. Notably, the HNEMD method establishes a mechanical analogue of the thermal transport process, and leverages linear response theory to calculate transport coefficients. This approach effectively eliminates finite-size effects and demonstrates high efficiency in lattice thermal conductivity calculations, with results closely aligning with experimental data. Here, we have performed three independent HNEMD simulations, each with a production time of 20 ns. Further details of the MD simulation are provided in Section 5 of Supporting Information. To calculate the effective system volume required for evaluating the three-dimensional effective thermal conductivity, it is necessary to define the thickness of the two-dimensional material. In this work, we define the thickness of monolayer $MoS_2$, $MoSe_2$, $MoTe_2$, $WS_2$, $WSe_2$, and $WTe_2$ as 6.2 Å, 6.6 Å, 7.2 Å, 6.2 Å, 6.6 Å, and 7.2 Å, respectively. We summarize the converged thermal conductivity values for the six monolayer and bilayer TMDs obtained from the HNEMD method in **Table S5** of Supporting Information Section 6, together with predicted values reported in the literature obtained using experiment and calculated methods. All the in-plane thermal conductivities of monolayer and bilayer TMDs are higher than current experiment results, as shown in **Fig. 6**a-c. These discrepancies between experimental measurements and our theoretical calculations can be attributed to several factors, including phonon boundary scattering due to the finite system length of the experimental samples, edge roughness, and defects and impurities present in the synthesized material[103, 104].

However, our results of monolayer and bilayer TMDs align with the higher values calculated using BTE method, following the decreasing order: $WS_2$ > $MoS_2$ > $WSe_2$ > $MoSe_2$ > $MoTe_2$ > $WTe_2$. Specially, at 300 K, the calculated in-plane thermal conductivities of monolayer TMDs using HNEMD method are 150 ± 6.2 Wm$^{-1}$K$^{-1}$ for $MoS_2$, 76.2 ± 4.3 Wm$^{-1}$K$^{-1}$ for $MoSe_2$, 41.7 ± 2.2 Wm$^{-1}$K$^{-1}$ for $MoTe_2$, 216.2 ± 7.7 Wm$^{-1}$K$^{-1}$ for $WS_2$, 76.7 ± 4.0 Wm$^{-1}$K$^{-1}$ for $WSe_2$, and 36.5 ± 3.1 Wm$^{-1}$K$^{-1}$ for $WTe_2$, respectively (see **Table 1**). The highest thermal conductivity of $WS_2$ is attributed to its wide phonon band gap, while the significant atomic mass disparity between W and S contributes to reduced phonon-phonon scattering, thereby enhancing its overall thermal conductivity[105]. Moreover, **Table 1** and **Fig. 6**d show that the calculated in-plane thermal conductivities of bilayer TMDs using the HNEMD method are 109 ± 1.6 Wm$^{-1}$K$^{-1}$ for $MoS_2$, 51.4 ± 2.2 Wm$^{-1}$K$^{-1}$ for $MoSe_2$, 28.0 ± 2.6 Wm$^{-1}$K$^{-1}$ for $MoTe_2$,

177.8 ± 8.8 Wm$^{-1}$K$^{-1}$ for WS$_2$, 64.2 ± 4.2 Wm$^{-1}$K$^{-1}$ for WSe$_2$, and 25.0 ± 1.0 Wm$^{-1}$K$^{-1}$ for WTe$_2$, respectively.

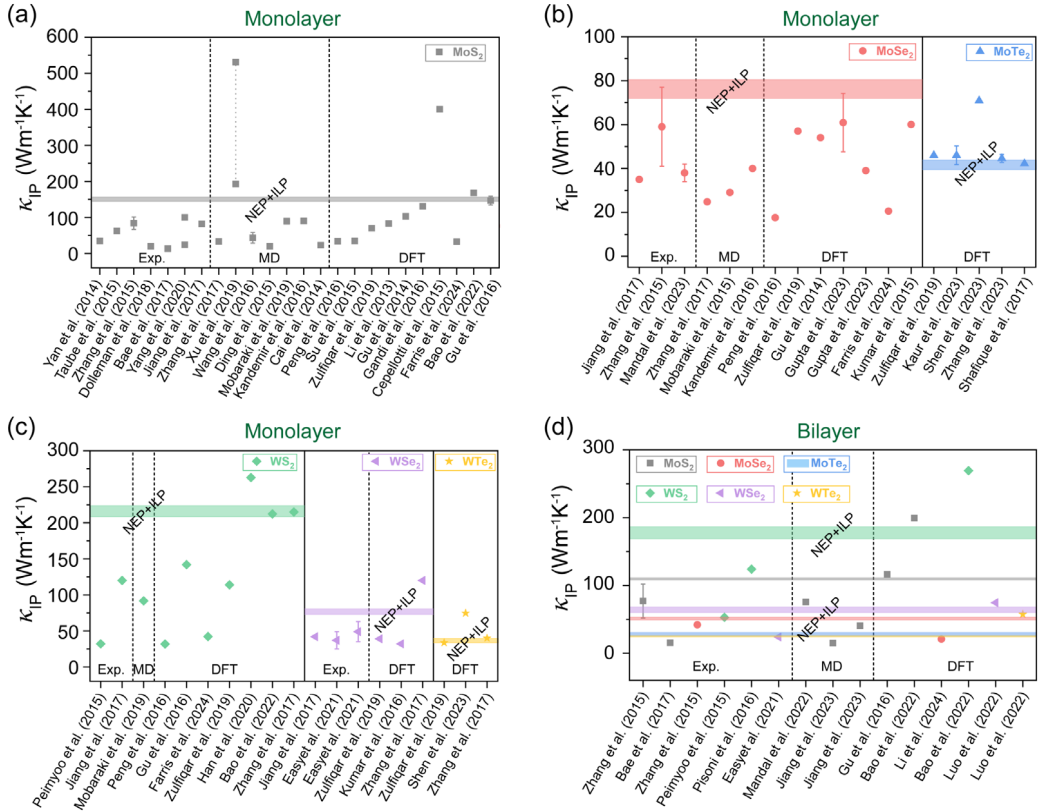

**Fig. 6 | In-plane thermal conductivity of TMD system.** Comparison of in-plane thermal conductivity of (a-c) monolayer and (d) bilayer TMDs (MoS$_2$, MoSe$_2$, MoTe$_2$, WS$_2$, WSe$_2$, and WTe$_2$) for different experiment and calculated methods[4, 103, 105-140]. More detailed reference data can be found in **Table S5** of Supporting Information Section 6.

*3.6. Cross-plane Thermal Conductivity*

To further demonstrate the applicability of the NEP+ILP force field for predicting the interlayer thermal transport in TMDs, we calculated the cross-plane thermal conductivity ($\kappa_{CP}$) of bulk MoS$_2$, MoSe$_2$, MoTe$_2$, WS$_2$, WSe$_2$, and WTe$_2$ at 300 K and zero pressure using the HNEMD method and further compared them to current DFT data and experimental results. Here, the simulations were conducted on a 20-layer square bulk system comprising a total of 17280 atoms, with each layer measuring 5 nm × 5 nm. More simulation details are consistent with the Section 5 of Supporting Information. **Table S6** in Section 6 of Supporting Information summarize the calculated cross-plane thermal conductivity obtained from the HNEMD method for the six TMDs, which are in excellent agreement with the available experiment and DFT values for bulk TMDs (see **Fig. 7**). Specially, the calculated $\kappa_{CP}$ has the following decreasing order: $WS_2$ > $MoS_2$ > $WSe_2$ > $MoSe_2$ > $WTe_2$ > $MoTe_2$. At 300 K, the $\kappa_{CP}$ of bulk TMDs using HNEMD method are $4.1 \pm 1.1$ Wm$^{-1}$K$^{-1}$ for $MoS_2$, $3.1 \pm 0.6$ Wm$^{-1}$K$^{-1}$ for $MoSe_2$, $3.0 \pm 1.1$ Wm$^{-1}$K$^{-1}$ for $MoTe_2$, $6.1 \pm 0.6$ Wm$^{-1}$K$^{-1}$ for $WS_2$, $3.9 \pm 0.6$ Wm$^{-1}$K$^{-1}$ for $WSe_2$ and $3.2 \pm 1.0$ Wm$^{-1}$K$^{-1}$ for $WTe_2$, respectively (see **Table 1**). These results support the validity of NEP+ILP computational framework and HNEMD method to study the cross-plane thermal transport of layered material interfaces.

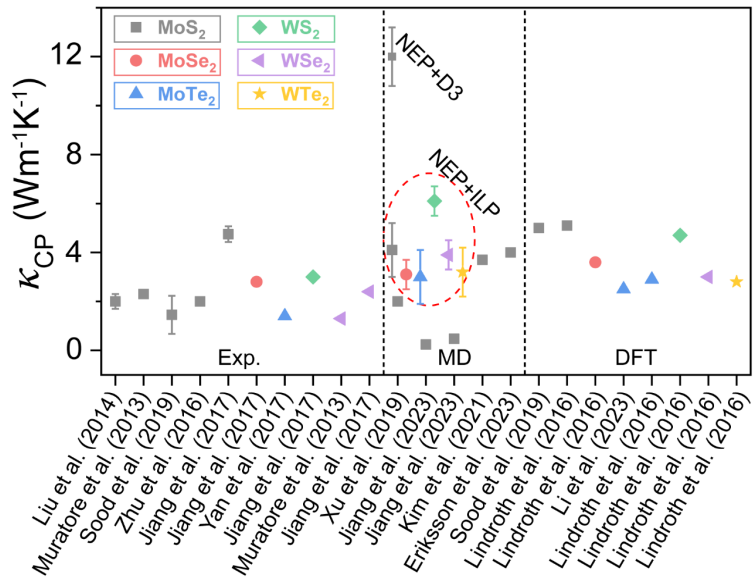

**Fig. 7 | Cross-plane thermal conductivity of TMD system.** Comparison of cross-plane thermal conductivity of bulk TMDs ($MoS_2$, $MoSe_2$, $MoTe_2$, $WS_2$, $WSe_2$, and $WTe_2$) for different experiment and calculated methods[4, 6, 27, 136, 141-147]. More detailed reference data can be found in **Table S6** of Supporting Information Section 6.

Table 1 | In-plane thermal conductivity $\kappa_{IP}$ and cross-plane thermal conductivity $\kappa_{CP}$ of TMDs ($MoS_2$, $MoSe_2$, $MoTe_2$, $WS_2$, $WSe_2$, and $WTe_2$) at room temperature, calculated using HNEMD bashed on NEP method for monolayer TMD system and NEP+ILP method for bilayer and bulk TMD system.

| TMD system | $\kappa_{IP}$ (Wm$^{-1}$K$^{-1}$) | | $\kappa_{CP}$ (Wm$^{-1}$K$^{-1}$) |
|---|---|---|---|
| | Monolayer | Bilayer | Bulk |
| $MoS_2$ | 150.4 ± 6.2 | 109.7 ± 1.6 | 4.1 ± 1.1 |
| $MoSe_2$ | 76.2 ± 4.3 | 51.4 ± 2.2 | 3.1 ± 0.6 |
| $MoTe_2$ | 41.7 ± 2.2 | 28.0 ± 2.6 | 3.0 ± 1.1 |
| $WS_2$ | 216.2 ± 7.7 | 177.8 ± 8.8 | 6.1 ± 0.6 |
| $WSe_2$ | 76.7 ± 4.0 | 64.2 ± 4.2 | 3.9 ± 0.6 |
| $WTe_2$ | 36.5 ± 3.1 | 25.0 ± 1.0 | 3.2 ± 1.0 |

## 4. COMPARISON BETWEEN NEP+ILP and NEP-D3 for $MoS_2$ SYSTEM

Recently, a different approach, NEP-D3, has been proposed to simultaneously model relatively short-ranged bonded interactions and relatively long-ranged dispersion interactions of layered materials[33]. To compare the NEP-D3 method and our NEP+ILP, we developed a NEP-D3 force field specifically for the $MoS_2$ system and performed mechanical and thermal transport MD simulations for $MoS_2$ systems, which are illustrated in **Fig. 8**. For the NEP-D3 dataset of the $MoS_2$ system, we incorporated bulk and bilayer configurations using the same strategy for monolayer systems. Moreover, configurations with different stacking types are included. The total dataset consists of 1099 frames for the NEP-D3 dataset of the $MoS_2$ system.

The NEP-D3 approach exhibits good performance in modeling twisted $MoS_2$ systems, as demonstrated through structural, vibrational, and thermal transport analyses. Atomic reconstruction simulations of twisted $MoS_2$ bilayers revealed moiré superlattice patterns—triangular in parallel and hexagonal in antiparallel configurations (**Fig. 8**a-b)—closely matching experimental and NEP+ILP-derived structures. Phonon dispersion calculations along the A→Γ→M→K→Γ pathway further aligned with experimental data (**Fig. 8**c), confirming the model's accuracy in capturing lattice dynamics. Thermal transport properties, evaluated using HNEMD simulations (see **Figure S19** in Section 5 of Supporting Information), showed that the $\kappa_{IP}$ for monolayer and bilayer $MoS_2$ (141.3 ± 4.7 Wm$^{-1}$K$^{-1}$ and 121.6 ± 1.0 Wm$^{-1}$K$^{-1}$, respectively) closely matched NEP+ILP results and aligned with BTE predictions. However, the $\kappa_{\mathrm{CP}}$ predictions (12.0 ± 1.2 Wm$^{-1}$K$^{-1}$) of bulk MoS$_2$ exceeded experimental values, which can be attributed to the underestimated equilibrium interlayer distance (6.0 Å) predicted by NEP-D3 method, in contrast to the larger value of 6.2 Å obtained from HSE+MBD-NL and NEP+ILP calculations.

Despite these successes in structural, vibrational, and thermal predictions, calculations of interlayer binding and sliding energies reveal a critical limitation: NEP-D3 significantly overestimates these values compared to HSE+MBD-NL and NEP+ILP calculations (**Fig. 8**d-i), indicating its inadequacy for accurately simulating interfacial mechanics and frictional behavior in layered TMD systems. It is worth noting that the sliding potential energy surface calculations are performed at the respective equilibrium interlayer distances predicted by each method. In contrast, NEP+ILP demonstrates good agreement with HSE+MBD-NL in interlayer energy calculations, further validating its suitability for modeling interfacial mechanics in TMD materials. The discrepancies between NEP+ILP and NEP-D3 results primarily stem from the differences in their underlying reference DFT methods used during model training (see Section 7 of Supporting Information). Additionally, the sliding PES predicted by NEP-D3 at different interlayer distances (6.0 Å and 6.2 Å) exhibits pronounced inconsistencies (see Section 7 of Supporting Information), primarily due to the absence of sliding configurations at the interlayer distance of 6.2 Å in the training dataset. As a result, NEP-D3 fails to accurately capture the interfacial potential energy distribution, underscoring its limited transferability. In comparison, NEP+ILP exhibits superior generalizability, making it a more reliable approach for studying interfacial interactions in layered materials (see Section 7 of Supporting Information).

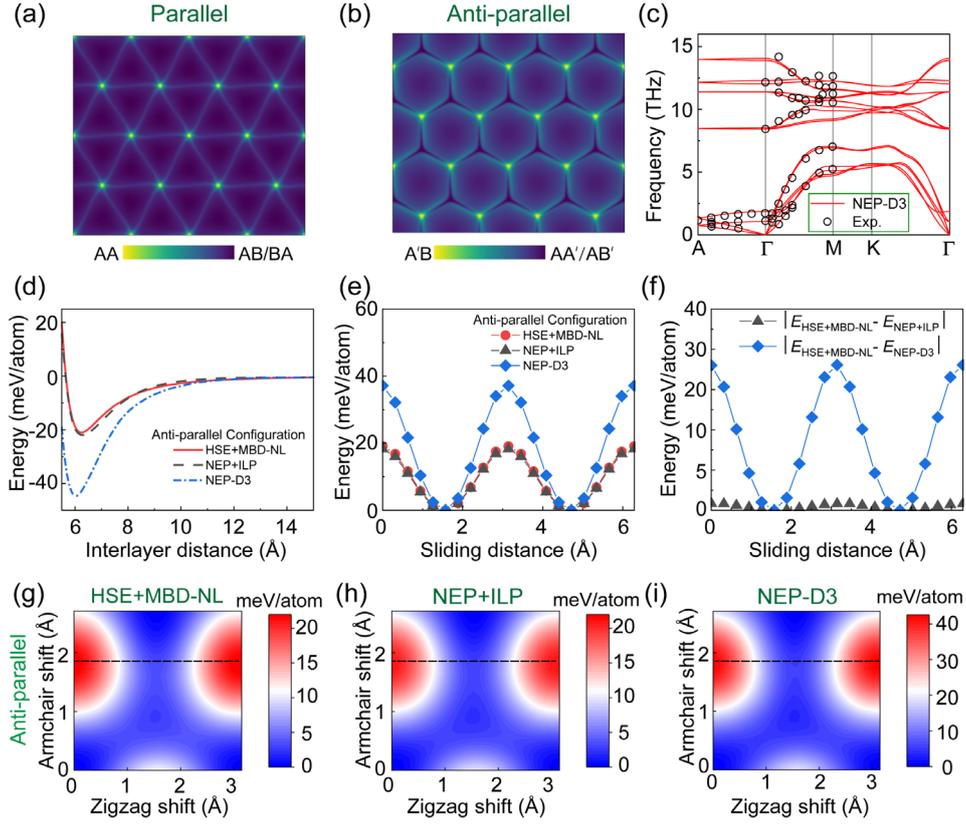

**Fig. 8 | Calculations of MoS₂ system using NEP-D3 method.** LRI[95, 96] corrugation of twisted MoS₂ bilayers in the parallel configuration (a) and the anti-parallel configuration (b) with $\theta$ = 0.25°, relaxed using the NEP-D3 force field. The yellow (black) markings stand for AA (AB and BA) stacks in parallel configuration and the AB′ (AA′ and A′B) stacks in anti-parallel configuration, respectively. (c) Phonon spectra of bulk MoS₂. Red solid lines are dispersion curves calculated using NEP-D3 force field. Black circles present the experimental results of bulk MoS₂[102]. (d) Interlayer binding energy (BE) of two AA′ stacking MoS₂ layers. Red solid lines, black dashed lines, and blue dashed lines are BE curves calculated using the HSE+MBD-NL, NEP+ILP, and NEP-D3, respectively. (e) Sliding energy of two AA′-stacked MoS₂ layers calculated using the HSE+MBD-NL (red circles), NEP+ILP (black triangles), and NEP-D3 (blue diomands), respectively. The considered lateral positions are indicated by the black dashed line in the unit cell, shown in panels (g-i). In panel (f), we presented the difference of sliding energy with $|E_{\text{HSE+MBD-NL}} - E_{\text{NEP+ILP}}|$ (balck triangles) and $|E_{\text{HSE+MBD-NL}} - E_{\text{NEP-D3}}|$ (blue diomands), respectively. Sliding energy surfaces of bilayer MoS₂ calculated at equilibrium interlayer distances using HSE+MBD-NL (g), NEP+ILP (h), and NEP-D3 (i), respectively. Specifically, the equilibrium interlayer distances of bilayer MoS₂ using HSE+MBD-NL, NEP+ILP, and NEP-D3 are 6.2, 6.2, and 6.0 Å, respectively. The reported energies are normalized by the total number of atoms in the cell. All the HSE+MBD-NL results are extracted from Ref. 24.

## 5. CONCLUSIONS

In summary, we have developed a hybrid computational framework that integrates an anisotropic interlayer potential with a machine learning potential to enable accurate molecular dynamics simulations for homogeneous TMD systems ($MoS_2$, $MoSe_2$, $MoTe_2$, $WS_2$, $WSe_2$, and $WTe_2$). The framework (NEP+ILP) yields bulk modulus values and in-plane thermal conductivity in good agreement with dispersion-augmented DFT predictions and cross-plane thermal conductivity aligned with experimental results. This demonstrates its utility as a comprehensive and efficient tool for investigating the mechanical, tribological, and thermal transport properties of TMD-based systems at large scales. The developed computational framework can be readily extended to other vdW heterostructures.

ASSOCIATED CONTENT

**Data Availability**

The registry-dependent interlayer potential for TMD system is available at [https://docs.lammps.org/pair_ilp_tmd.html](https://docs.lammps.org/pair_ilp_tmd.html). The NEP interface to LAMMPS is available at [https://github.com/brucefan1983/NEP_CPU](https://github.com/brucefan1983/NEP_CPU). In GPUMD, documentation for the hybrid NEP+ILP potential is available at [https://gpumd.org/dev/potentials/nep_ilp.html](https://gpumd.org/dev/potentials/nep_ilp.html). The source codes of hybrid NEP+ILP force field is currently available in the master branch of GPUMD ([https://github.com/brucefan1983/GPUMD](https://github.com/brucefan1983/GPUMD)) and will be released later with GPUMD-v4.0.

**Supporting Information**

Supporting Information contains the following sections: Details of NEP Model Training, Details of Interlayer Potential Parameterization, Fitting for Bulk Modulus, Interlayer Registry Index of Homogeneous $MoTe_2$ and $WTe_2$, HNEMD Simulations, In-plane and Cross-plane Thermal Conductivity of TMDs, and Binding and Sliding Energy of Bilayer $MoS_2$ Calculated using Different DFT Methods.

**Corresponding Author**

*E-mail: [w.g.ouyang@whu.edu.cn](mailto:w.g.ouyang@whu.edu.cn).

**Notes**

The authors declare no competing financial interest.

**Acknowledgments**

W.O. acknowledges support from the National Natural Science Foundation of China (Nos. 12472099 and U2441207) and the Fundamental Research Funds for the Central Universities (Nos. 2042025kf0050 and 600460100). T.L. and J.X. acknowledge the National Natural Science Foundation of China (Grant No. U20A20301) and the Research Grants Council of Hong Kong (Grant No. AoE/P-701/20). Computations were conducted at the Supercomputing Center of Wuhan University, the National Supercomputer TianHe-1(A) Center in Tianjin and Computing Center in Xi'an.

**Author Contributions**

#W.J., H.B., and T.L. contributed equally to this work. W.O. conceived the research and supervised the project. W.J. participated in the construction of the NEP model and the registry-dependent ILP, and conducted all DFT and MD simulations. H.B. implemented the NEP+ILP framework in LAMMPS and GPUMD. T.L. contributed to the training of the NEP model and assisted in the DFT calculations. All authors contributed to data analysis and participated in writing the manuscript.